\documentclass[showpacs,amsmath,amssymb]{revtex4}
\usepackage{graphicx}
\usepackage{dcolumn}
\usepackage{bm}
\newcommand{\be}{\begin{eqnarray}}
\newcommand{\en}{\end{eqnarray}}
\newcommand{\ben}{\begin{eqnarray*}}
\newcommand{\enn}{\end{eqnarray*}}
\newcommand{\pa}{\partial}

\newcommand{\f}{\frac}

\newcommand{\p}{\paragraph{}}
\newcommand{\bi}{\begin{itemize}}
\newcommand{\ei}{\end{itemize}}

\newcommand{\la}{\langle}
\newcommand{\ra}{\rangle}

\renewcommand{\r}{\rho}
\renewcommand{\p}{\bot}
\renewcommand{\a}{\alpha}
\renewcommand{\b}{\beta}

\begin{document}
\title{A critical note on one-eighth law in 2D turbulence}
\author{Sagar Chakraborty}
\email{sagar@bose.res.in}
\affiliation{S.N. Bose National Centre for Basic Sciences\\Saltlake, Kolkata 700098, India}
\date{12 June, 2008}
\begin{abstract}
We critically revisit the various attempts to prove one eighth law in 2D turbulence and reconcile them.
Herein, the one-eight law has been proved for unforced 2D incompressible high Reynolds number turbulence.
An exact expression of the time derivative of two-point second order velocity correlation function is also derived for the enstrophy cascade dominated regime.
\end{abstract}
\maketitle
%
The phenomenological model of Kolmogorov\cite{Kolmogorov} remains the cornerstone of our knowledge about the theory of 3D incompressible high Reynolds number turbulence.
The most important non-trivial result that it yielded is the four-fifths law:
\ben
S_3\equiv\left<\left[\left\{\vec{v}(\vec{r}+\vec{l})-\vec{v}(\vec{r})\right\}.\f{\vec{l}}{|\vec{l}|}\right]^3\right>=-\f{4}{5}\varepsilon l
\enn
where $\vec{v}$ is the velocity field and $\varepsilon$ is the rate per unit mass at which energy is being transferred through the inertial range.
This rare exact result has stood the rigorous tests put forward by experiments and numerics.
Theoretically this result could be derived in two ways\cite{Bhattacharjee}: i) the original Kolmogorov method explained in details in the fluid dynamics text due to Landau and Lifshitz\cite{Landau}.
(There is no external forcing in this approach and the equality of dissipation rate and forcing rate for the energy is never enforced), and ii) a field-theoretic method. (Here an external forcing maintains a steady state turbulence).
The field theoretic technique invokes the so-called ``dissipation anomaly" for deriving the law.
\\
Even in 2D turbulence one would like to seek such a classic law concerning two-point third order structure function $S_3$ using one or the other of the above-mentioned methods.
Recent attempts to do so by Lindborg\cite{Lindborg2D}, Bernard\cite{Bernard} and Chakraborty\cite{Sagar} amounted to some revealing results.
The equivalence between the two techniques has also got highlighted by virtue of these results.
However, one point of confusion has remained.
It concerns the inability to derive one-eighth law in the enstrophy cascade dominated regime in unforced 2D turbulence using Kolmogorov-Landau method\cite{Sagar}.
It may be contrasted with the fact that the one-eight law seems to be correctly derived in 2D forced turbulence\cite{Lindborg2D,Bernard}.
\\
Frisch\cite{Frisch} has observed that as certain mathematical steps leading to the proof of one-eighth law in forced 2D turbulence are of purely kinematic nature and consequently they should also lead to the one-eighth law in unforced version of 2D turbulence.
If such is the case then it would modify certain conclusions of reference\cite{Sagar} and would extend the validity of one-eighth law into the regime of unforced 2D turbulence.
We shall see shortly in what follows that this indeed is the case.
\\ 
It is a well-established fact that there exists a direct-cascade of enstrophy in two-dimensional (2D) turbulence.
One defines total enstrophy as $\Gamma=\f{1}{2}\int_{\textrm{all space}}\omega^2 d^2\vec{\r}$ where $\omega=\pa_xv_y-\pa_yv_x$ is the vorticity in the Cartesian coordinates; $\vec{v}$ being the velocity field.
As we shall consider incompressible fluids only ($\vec{\nabla}.\vec{v}=0$), we shall take the density to be unity and let $\vec{\r}$ take over the task of representing position vector in 2D plane.
The enstrophy flows through the inertial range and gets dissipated near the dissipation scale.
Using the antisymmetric symbol $\varepsilon_{\a\b}$ that has four components, viz. $\varepsilon_{11}=\varepsilon_{22}=0$ and $\varepsilon_{12}=-\varepsilon_{21}=1$, one may define the mean rate of dissipation of enstrophy per unit mass as:
\be
&{}&\eta\equiv\nu\la \vec{\nabla}\omega.\vec{\nabla}\omega\ra\nonumber\\
\Rightarrow&{}&\eta=\nu\varepsilon_{\tau\a}\varepsilon_{\theta\b}\la (\pa_\tau\pa_\gamma v_{\a})(\pa_\theta\pa_\gamma v_{\b})\ra
\label{0}
\en
Here, angular brackets denote an averaging procedure which averages over all possible positions of points $1$ and $2$ at a given instant of time and a given separation.
Now, if $\vec{v}_1$ and $\vec{v}_2$ represent the fluid velocities at the two neighbouring points at $\r_1$ and $\r_2$ respectively, one may define rank two correlation tensor:
\be
B_{\a\b}\equiv\la(v_{2\a}-v_{1\a})(v_{2\b}-v_{1\b})\ra
\label{1}
\en
For simplicity, we shall take a rather idealised situation of turbulence flow which is homogeneous and isotropic on every scale, a case achievable in practice in a vigorously-shaken-fluid left to itself.
The component of the correlation tensor will obviously, then, be dependent on time, a fact which won't be shown explicitly in what follows.
As the features of local turbulence is independent of averaged flow, the result derived below is applicable also to the local turbulence at distance $\r$ much smaller than the fundamental scale.
Isotropy and homogeneity suggests the following general form for $B_{\a\b}$
\be
B_{\a\b}=A_1(\r)\delta_{\a\b}+A_2(\r)\r^o_\a\r^o_\b
\label{2}
\en
where $A_1$ and $A_2$ are functions of time and $\r$.
The Greek subscripts can take two values $\r$ and $\p$ which respectively mean the component along the radial vector $\r$ and the component in the transverse direction.
Einstein's summation convention will be used extensively.
Also,
\ben
\vec{\r}=\vec{\r}_2-\vec{\r}_1,\phantom{xxx}\r^o_\a\equiv\r_\a/{|\vec{\r}|},\phantom{xxx}\r^o_\r=1,\phantom{xxx}\r^o_\p=0
\enn
using which in the relation (\ref{2}), one gets:
\be
B_{\a\b}=B_{\p\p}(\delta_{\a\b}-\r^o_\a\r^o_\b)+B_{\r\r}\r^o_\a\r^o_\b
\label{3}
\en
One may break the relation (\ref{1}) as
\be
B_{\a\b}=\la v_{1\a}v_{1\b}\ra+\la v_{2\a}v_{2\b}\ra- \la v_{1\a}v_{2\b}\ra-\la v_{2\a}v_{1\b}\ra
\label{4}
\en
and defining $b_{\a\b}\equiv\la v_{1\a}v_{2\b}\ra$ one may proceed, keeping in mind the isotropy and the homogeneity, to write
\be
B_{\a\b}=\la v^2\ra\delta_{\a\b}-2b_{\a\b}
\label{6}
\en
\noindent Let's again define: $b_{\a\b,\gamma}\equiv\la v_{1\a}v_{1\b}v_{2\gamma}\ra$.
Invoking homogeneity and isotropy once again along with the symmetry in the first pair of indices, one may write the most general form of the third rank Cartesian tensor for this case as
\be
b_{\a\b,\gamma}&=&C(\r)\delta_{\a\b}\r^o_\gamma+D(\r)(\delta_{\gamma\b}\r^o_\a+\delta_{\a\gamma}\r^o_\b)+F(\r)\r^o_\a\r^o_\b\r^o_\gamma
\label{10}
\en
where, $C$, $D$ and $F$ are functions of $\r$.
Incompressibility dictates:
\be
\f{\pa}{\pa\rho_{2\gamma}}b_{\a\b,\gamma}=\f{\pa}{\pa\rho_{\gamma}}b_{\a\b,\gamma}=0\\
\Rightarrow C'\delta_{\a\b}+\f{C}{\r}\delta_{\a\b}+\f{2D}{\r}\delta_{\a\b}+\f{2D'}{\r^2}\r_\a\r_\b-\f{2D}{\r^3}\r_\a\r_\b+\f{F'}{\r^2}\r_\a\r_\b+\f{F}{\r^3}\r_\a\r_\b=0
\label{11}
\en
Here prime ($'$) denotes derivative w.r.t. $\rho$.
Putting $\a=\b$ in equation (\ref{11}) one gets:
\be
2C+2D+F=\f{\textrm{constant}}{\r}=0
\label{11-12}
\en
where, it as been imposed that $b_{\a\b,\gamma}$ should remain finite for $\r=0$.
Again, using equation (\ref{11}), putting $\a\ne\b$ and manipulating a bit one gets:
\be
D=-\f{1}{2}(\r C'+C)
\label{12}
\en
using which in relation (\ref{11-12}), one arrives at the following expression for $F$:
\be
F=\r C'-C
\label{13}
\en
Defining
\be
B_{\a\b\gamma}&\equiv&\la(v_{2\a}-v_{1\a})(v_{2\b}-v_{1\b})(v_{2\gamma}-v_{1\gamma})\ra\\
&=&2(b_{\a\b,\gamma}+b_{\gamma\b,\a}+b_{\a\gamma,\b})
\label{13-14}
\en
and putting relations (\ref{12}) and (\ref{13}) in equation (\ref{13-14}) and using relation (\ref{10}), one gets:
\be
&{}&B_{\a\b\gamma}=-2\r C'(\delta_{\a\b}\r^o_\gamma+\delta_{\gamma\b}\r^o_\a+\delta_{\a\gamma}\r^o_\b)+6(\r C'-C)\r^o_\a\r^o_\b\r^o_\gamma\\
\label{14}
\Rightarrow&{}&S_3\equiv B_{\r\r\r}=-6C
\label{15}
\en
which along with relations (\ref{12}), (\ref{13}) and (\ref{10}) yields the following expression:
\be
&{}&b_{\a\b,\gamma}=-\f{S_3}{6}\delta_{\a\b}\r^o_\gamma+\f{1}{12}(\r S'_3+S_3)(\delta_{\gamma\b}\r^o_\a+\delta_{\a\gamma}\r^o_\b)-\f{1}{6}(\r S'_3-S_3)\r^o_\a\r^o_\b\r^o_\gamma
\label{16}
\en
Navier-Stokes equation gives:
\be
\f{\pa}{\pa t}v_{1\a}=-v_{1\gamma}\pa_{1\gamma}v_{1\a}-\pa_{1\a}p_1+\nu\pa_{1\gamma}\pa_{1\gamma}v_{1\a}
\label{17}\\
\f{\pa}{\pa t}v_{2\b}=-v_{2\gamma}\pa_{2\gamma}v_{2\b}-\pa_{2\b}p_2+\nu\pa_{2\gamma}\pa_{2\gamma}v_{2\b}
\label{18}
\en
Multiplying equations (\ref{17}) and (\ref{18}) with $v_{2\b}$ and $v_{1\a}$ respectively and adding subsequently, one gets the following:
\be
\f{\pa}{\pa t}\la v_{1\a}v_{2\b}\ra&=&-\pa_{1\gamma}\la v_{1\gamma}v_{1\a}v_{2\b}\ra-\pa_{2\gamma}\la v_{2\gamma}v_{1\a}v_{2\b}\ra\nonumber\\
&{}&-\pa_{1\a}\la p_1v_{2\b}\ra-\pa_{2\b}\la p_2v_{1\a}\ra+\nu \pa_{1\gamma}\pa_{1\gamma}\la v_{1\a}v_{2\b}\ra\nonumber\\
&{}&+\nu \pa_{2\gamma}\pa_{2\gamma}\la v_{1\a}v_{2\b}\ra
\label{19}
\en
Due to isotropy, the correlation function for the pressure and velocity ($\la p_1\vec{v}_2\ra$) should have the form $f(\r)\vec{\r}/|\vec{\r}|$.
But since $\pa_{\a}\la p_1{v}_{2\a}\ra=0$ due to solenoidal velocity field, $f(\r)\vec{\r}/|\vec{\r}|$ must have the form $\textrm{constant}\times(\vec{\r}/|\vec{\r}|^2)$ that in turn must vanish to keep correlation functions finite even at $\r=0$.
Thus, equation (\ref{19}) can be written as:
\be
\f{\pa}{\pa t}b_{\a\b}=\pa_{\gamma}(b_{\a\gamma,\b}+b_{\b\gamma,\a})+2\nu \pa_\gamma\pa_\gamma b_{\a\b}
\label{20}
\en
For isotropic and homogeneous turbulence, the condition of incompressibility gives the easily derivable well-known result:
\be
4\pa_\gamma b_{\a\gamma,\a}=\pa_\gamma B_{\a\a\gamma}
\label{neww1}
\en
Defining $W\equiv\la\omega_1\omega_2\ra$ and noting that $W=-\pa_\delta\pa_\delta b_{\a\a}$, we get from relations (\ref{20}) and (\ref{neww1}):
\be
-\f{\pa W}{\pa t}=\f{1}{2}\pa_\delta\pa_\delta(\pa_{\gamma}B_{\a\a\gamma})-2\nu\pa_\delta\pa_\delta W
\label{neww2}
\en
Again, if one defines $\Omega\equiv\la(\omega_2-\omega_1)(\omega_2-\omega_1)\ra$, for homogeneous isotropic turbulence one may write $\Omega=2\la\omega^2\ra-2W$.
So, equation (\ref{neww2}) can be manipulated into the following:
\be
&&\f{1}{2}\f{\pa \Omega}{\pa t}-\f{\pa \la\omega^2\ra}{\pa t}=\f{1}{2}\pa_\delta\pa_\delta\pa_{\gamma}B_{\a\a\gamma}+\nu\pa_\delta\pa_\delta \Omega-2\nu\pa_\delta\pa_\delta\la\omega^2\ra\\
\Rightarrow&&\pa_\delta\pa_\delta\pa_{\gamma}B_{\a\a\gamma}=4\eta\\
\Rightarrow&&B_{\a\a\gamma}=\f{1}{4}\eta\rho^3\label{neww3}
\en
Here, we have assumed $\f{\pa \Omega}{\pa t}$ to be relatively negligible and let $\nu\rightarrow 0$ so that the terms proportional to $\nu$ vanish.
Also, we have recalled that $\f{1}{2}\f{\pa \la\omega^2\ra}{\pa t}=-\eta$.
From equations (\ref{10}) and (\ref{13-14}), and the condition of incompressibility, it readily follows that $B_{\bot\bot\rho}=\f{\rho}{3}\f{\pa}{\pa\rho}B_{\r\r\r}$ putting which in expression (\ref{neww3}) and integrating subsequently (keeping in mind that $B_{\r\r\r}$ shouldn't blow up at $\rho=0$), we arrive at:
\be
B_{\r\r\r}=+\f{1}{8}\eta\r^3\label{neww4}
\en
This is the one-eighth law for the unforced 2D incompressible turbulence proved using the Kolmogorov-Landau approach.
\\
Let us go back to equation (\ref{20}).
Using expressions (\ref{6}) and (\ref{16}), one can rewrite the equation as:
\be
&{}&\f{1}{2}\f{\pa}{\pa t}\la v^2\ra-\f{1}{2}\f{\pa}{\pa t}B_{\r\r}=\nu \pa_\gamma\pa_\gamma\la v^2\ra+\f{1}{6\r^3}\f{\pa}{\pa \r}\left(\r^3B_{\r\r\r}\right)-\f{\nu}{\r}\f{\pa}{\pa \r}\left(\r\f{\pa B_{\r\r}}{\pa \r}\right)
\label{21}
\en
As we are interested in the enstrophy cascade, the first term in the R.H.S. is zero due to homogeneity and the first term in the L.H.S. is zero because of energy remains conserved in 2D turbulence in the inviscid limit (and it is the high Reynolds number regime that we are interested in); it cannot be dissipated at smaller scales.
Also, as we are interested in the forward cascade which is dominated by enstrophy cascade, on the dimensional grounds in the inertial region $B_{\r\r}$ (as it may depend only on $\eta$ and $\r$) may be written as:
\be
\f{\pa}{\pa t}B_{\r\r}=A\eta\r^2
\label{22}
\en
where $A$ is a numerical proportionality constant.
Hence, using the relation (\ref{22}), the equation (\ref{21}) reduces to the following differential equation:
\be
\f{1}{6\r^3}\f{\pa}{\pa \r}\left(\r^3B_{\r\r\r}\right)=\f{\nu}{\r}\f{\pa}{\pa \r}\left(\r\f{\pa B_{\r\r}}{\pa \r}\right)-\f{A}{2}\eta\r^2
\label{23}
\en
which when solved using relation(\ref{15}) in the limit of infinite Reynolds number ($\nu\rightarrow 0$), one gets
\be
B_{\r\r\r}=-\f{A\eta}{2}\r^3
\label{24}
\en
Comparing it with the expression (\ref{neww4}) for the two-point third order velocity correlation function for the isotropic and homogeneous 2D decaying turbulence in the inertial range of the forward cascade, we determine the value of $A$ to be -1/4.
Therefore, relation (\ref{22}) yields
\be
\f{\pa}{\pa t}B_{\r\r}=-\f{1}{4}\eta\r^2
\label{25}
\en
This, to the best of our knowledge, is yet another exact new result that has to be verified experimentally and numerically to test its validity.
\\
Having answered the confusion mentioned earlier regarding the one-eighth law, we turn critical to yet another issue that was not addressed in any of the references mentioned in the beginning.
We have been careless enough to assume the existence of $\eta$ when $\nu\rightarrow 0$.
If $\eta$ doesn't exist, the one-eighth law is yet again in jeopardy.
It is really unfortunate for the law that it has been rigourously proved\cite{Eyink2,Tran} that enstrophy dissipation is not possible for any 2D Euler solutions with finite enstrophy.
Thus, $\eta$ may exist in the inviscid limit only when one takes rather ill-defined initial conditions for which the total initial enstrophy is infinite.
In view of this, one must not be surprised at all if numerics and experiments fail to uphold the one-eighth law in many a situation.
The law of 2D turbulence, therefore, doesn't enjoy the same classic status as the Kolmogorov law of 3D turbulence.
\\
\\
The author would like to acknowledge his friends Prasad and Samriddhi, and his supervisor Prof. J. K. Bhattacharjee for the fruitful discussions. Also, CSIR (India) is gratefully acknowledged for awarding fellowship to the author. The author is indebted to Prof. Uriel Frisch and Prof. E. Lindborg for helpful correspondences.


\begin{thebibliography}{99}
%
\bibitem{Kolmogorov} A. N. Kolmogorov, {\it Dissipation of the energy in the locally isotropic turbulence}, {Dokl. Akad. Nauk SSSR, {\bf{32}}}, 1 (1941); (English translation: {Proc. R. Soc. Lond. A {\bf{434}}}, 15 (1991)).
%
\bibitem{Bhattacharjee}  J. K. Bhattacharjee and S. Bhattacharyya, {\it Non-linear Dynamics Near and Far from Equilibrium}, (Hindustan Book Agency, New Delhi), (1995).
%
\bibitem{Landau}  L. D. Landau and E. M. Lifshitz, {\it Fluid Mechanics, Second Edition: Volume 6 (Course of Theoretical Physics)}, (Reed Educational and Professional Publishing Ltd), (1987).
%
\bibitem{Lindborg2D} E. Lindborg, {\it Can the atmospheric kinetic energy spectrum be explained by two-dimensional turbulence}, { J. Fluid Mech. {\bf {388}}}, 259 (1999).
%
\bibitem{Bernard} D. Bernard, {\it Three-point velocity correlation functions in two-dimensional forced turbulence}, {Phys. Rev. E {\bf{60}}}, 6184 (1999).
%
\bibitem{Sagar} S. Chakraborty, {\it On the use of the Kolmogorov-Landau approach in deriving various correlation functions in two-dimensional incompressible turbulence}, { Phys. Fluids {\bf {19}}}, 085110 (2007).
%
\bibitem{Frisch} U. Frisch, Private communication.
%
\bibitem{Eyink2} G. L. Eyink, {\it Dissipation in turbulent solutions of 2D Euler equations}, {Nonlinearity {\bf{14}}}, 787 (2001).
%
\bibitem{Tran} C. V. Tran and D. G. Dritschel, {\it Vanishing enstrophy dissipation in two-dimensional Navier Stokes turbulence in the inviscid limit}, { J. Fluid Mech. {\bf {559}}}, 107 (2006).
%
\end{thebibliography}
\end{document}